\documentclass[10pt,aps,prl,superscriptaddress, nobibnotes]{revtex4-2}
\usepackage{amsmath,amsfonts,amssymb,verbatim,mathtools}
\usepackage{amsfonts}
\usepackage{graphicx}
\usepackage[normalem]{ulem}
\usepackage{color}

\usepackage{dcolumn}
\usepackage{tabularx}
\usepackage{keyval}
\usepackage{multirow}

\newcommand{\QMI}{Quantum Matter Institute, University of British Columbia, Vancouver, British Columbia, V6T 1Z4, Canada}

\newcommand{\UBC}{Department of Physics $\&$ Astronomy, University of British Columbia, Vancouver, British Columbia, V6T 1Z1, Canada.}

\newcommand{\CLS}{Canadian Light Source, Saskatoon, Saskatchewan, S7N 2V3, Canada.}

\newcommand{\MPI}{Max Planck Institute for Solid State Research, Stuttgart, Germany.}

\newcommand{\AIST}{National Institute of Advanced Industrial Science and Technology (AIST), Tsukuba, Ibaraki 305-8568, Japan}

\newcommand{\SSRL}{Stanford
Synchrotron Radiation Light Source, SLAC National Accelerator Laboratory, Menlo
Park, CA 94025, USA}

\newcommand{\sissa}{International School for Advanced Studies (SISSA), Via Bonomea 265, 34136 Trieste, Italy}

\newcommand{\CNR}{Istituto Officina dei Materiali (CNR-IOM), Via Bonomea 265, 34136 Trieste, Italy}

\begin{document}

\title{Coherence and pairing enhancement in a trilayer cuprate \\with layer-selective charge order}

\author{Steef Smit \textsuperscript{\textdagger}}
\thanks{These authors contributed equally to this work.}
\affiliation{\QMI}
\affiliation{\UBC}

\author{Martin Bluschke}
\thanks{These authors contributed equally to this work.}
\affiliation{\QMI}
\affiliation{\UBC}

\author{Peter Moen}
\affiliation{\QMI}
\affiliation{\UBC}

\author{Niclas Heinsdorf}
\affiliation{\QMI}
\affiliation{\UBC}

\author{Edoardo Zavatti}
\affiliation{\sissa}

\author{Gabriele Bellomia}
\affiliation{\sissa}

\author{Samuele Giuli}
\affiliation{\sissa}

\author{Sydney K.Y. Dufresne}
\affiliation{\QMI}
\affiliation{\UBC}

\author{Cissy T. Suen}
\affiliation{\QMI}
\affiliation{\UBC}
\affiliation{\MPI}

\author{Valentin Zimmermann}
\affiliation{\QMI}
\affiliation{\UBC}
\affiliation{\MPI}

\author{Christine Au-Yeung}
\affiliation{\QMI}
\affiliation{\UBC}

\author{Sergey Zhdanovich}
\affiliation{\QMI}
\affiliation{\UBC}

\author{Jerry I. Dadap}
\affiliation{\QMI}
\affiliation{\UBC}

\author{Pinder Dosanjh}
\affiliation{\QMI}
\affiliation{\UBC}

\author{Marta Zonno}
\affiliation{\CLS}

\author{Sergey Gorovikov}
\affiliation{\CLS}

\author{Heemin Lee}
\affiliation{\SSRL}

\author{Cheng-Tai Kuo}
\affiliation{\SSRL}

\author{Jun-Sik Lee}
\affiliation{\SSRL}

\author{Dongjoon Song}
\affiliation{\QMI}
\affiliation{\UBC}

\author{Shigeyuki Ishida}
\affiliation{\AIST}

\author{Hiroshi Eisaki}
\affiliation{\AIST}

\author{Bernhard Keimer}
\affiliation{\MPI}

\author{Matteo Michiardi}
\affiliation{\QMI}
\affiliation{\UBC}

\author{Ilya S. Elfimov}
\affiliation{\QMI}
\affiliation{\UBC}

\author{Giorgio Levy}
\affiliation{\QMI}
\affiliation{\UBC}

\author{David J. Jones}
\affiliation{\QMI}
\affiliation{\UBC}

\author{Massimo Capone}
\affiliation{\sissa}
\affiliation{\CNR}

\author{Andrea Damascelli}
\email{Corresponding authors: Steef Smit: steef.smit@ubc.ca; Andrea Damascelli: damascelli@physics.ubc.ca}
\affiliation{\QMI}
\affiliation{\UBC}

\begin{abstract} 
{\bf Trilayer cuprates hold the record for the highest superconducting critical temperatures ($T_{\text{c}}$), yet the underlying mechanism remains elusive. Using time- and angle-resolved photoemission spectroscopy (tr-ARPES), we uncover a striking interplay between charge order, superconducting gap magnitude, and quasiparticle coherence in Bi$_2$Sr$_2$Ca$_2$Cu$_3$O$_{10+\delta}$ (Bi2223). We present ARPES-based evidence of charge order on the inner CuO$_2$ plane -- as confirmed via resonant x-ray scattering (RXS) -- and show that the same inner plane hosts a large superconducting gap matching its doping, firmly establishing it as underdoped. Despite its underdoped nature, the inner plane also exhibits an exceptional degree of quasiparticle coherence; suppressing charge-order fluctuations further enhances this, making it comparable to that of the overdoped outer planes at elevated electronic temperatures. At the same time, on the overdoped outer planes we find larger superconducting gaps than what is seen on comparably doped single and bilayer cuprates, indicating an unexpected and simultaneous enhancement of both coherence and pairing interactions in Bi2223. These findings, supported by complementary single- and three-layer Hubbard model calculations, reveal a unique interlayer mechanism in which both pairing strength and phase coherence are optimized when interfacing planes with distinct hole concentrations, providing new microscopic insight into the record $T_{\text{c}}$ of Bi2223.}
\end{abstract}

\maketitle

\subsection{Introduction}

The phase diagram of the unconventional cuprate superconductors is characterized by a complex interplay of electronic phases whose origin and mutual relationships remain only partially understood~\cite{Keimer2015}. Systematic investigations have achieved consensus that while the strongest pairing interactions -- as indicated by spectral gap amplitudes -- are present in the underdoped region close to the strongly correlated Mott insulating state~\cite{Hashimoto2014}, the excitations here are generally much less coherent than those found in the optimally and overdoped regimes~\cite{Feng2000,Fournier2010}. The opposing trends in the doping dependent energy scales associated with pairing and phase coherence have been proposed to shape the superconducting dome~\cite{Emery1995}. Notably, it is the trilayer cuprates that reach the highest known superconducting critical temperatures ($T_{\text{c}}$) under ambient conditions. They contain multiple CuO$_{2}$ planes with different hole concentrations within each unit cell~\cite{Trokiner1991,Kotegawa2001}, with the inner plane (IP) and outer planes (OP) of the trilayer block exhibiting markedly different Fermi surface (FS) volumes. This raises the question whether these differently doped planes strictly adhere to the distinct phenomenologies observed across the cuprate phase diagram, or if some form of interlayer coupling between the planes is responsible for the high $T_{\text{c}}$. \\
When coupled to one another, the planes in these multilayer stacks may give rise to electronic properties that are distinct from the sum of the individual layers~\cite{Kivelson2002, Berg2008}. For example, in the 3- and 5-layer cuprates HgBa$_2$Ca$_2$Cu$_3$O$_{8+\delta}$ (Hg1223)~\cite{Oliviero2022} and Ba$_2$Ca$_4$Cu$_5$O$_{10}$(F,O)$_2$~\cite{Kunisada2020}, the existence of a FS in the presence of antiferromagnetic (AFM) order may be responsible for the elevated superconducting critical temperature.  In order to understand how interlayer interactions within the multilayer structure can enhance $T_{\text{c}}$, it is necessary to disentangle the electronic contributions of each layer. Here we perform time- and angle-resolved photoemission spectroscopy (tr-ARPES)~\cite{Boschini2024a} on trilayer Bi$_2$Sr$_2$Ca$_2$Cu$_3$O$_{10+\delta}$ (Bi2223) to selectively evaluate the single-particle spectral function associated with the inner and outer planes of the trilayer. Our measurements reveal a striking dichotomy in the pump-induced renormalization of spectral weight of the distinct underdoped ($p$ = 0.08) and overdoped ($p$ = 0.25) FS, originating from the inner and outer CuO$_2$ planes, respectively. By exploring the momentum and binding-energy dependence of the spectral weight response, we conclude that the observed dynamics are due to the pump-induced melting of an incipient $Q \sim $ 0.33 r.l.u. short-range charge order (CO) that is confined to the inner of the three CuO$_{2}$ planes. Although fluctuating CO correlations have been observed in all major families of cuprate superconductors~\cite{Keimer2015, Comin2016a, Frano2020}, clear signatures have remained elusive in ARPES, likely due to the extremely short correlation lengths. The sensitivity of non-equilibrium electronic dynamics to the presence of CO fluctuations allows the tr-ARPES measurements to detect these short-range correlations. The coexistence of CO -- as also confirmed via complementary resonant x-ray scattering (RXS) -- with a significantly larger superconducting gap on the IP aligns with the familiar phenomenology of underdoped cuprates, firmly establishing the underdoped nature of the IP. However, in stark contrast to expectations and in particular to the behavior typically observed on the underdoped single and bilayer cuprates, the IP also exhibits an extraordinary degree of quasiparticle (QP) coherence, as evidenced by its QP weight and lifetime. Moreover, quenching the fluctuating CO in our tr-ARPES experiments further enhances the inner-plane QP weight, making it comparable to that of the overdoped outer planes at elevated electronic temperatures. The superconducting gap on the OP, albeit smaller in magnitude than on the IP, is still significantly larger than what is seen on comparably doped Bi2201 and Bi2212. By comparing these results to theoretical calculations of a three-layer Hubbard model, we propose that the unexpected QP weight and gap magnitude of Bi2223 arise from interlayer interactions between the inner and outer planes. The denser charge fluid of the OP enhances QP weight on the strongly correlated IP, and the stronger pairing of the IP is shared with the OP. Such a layer-coupling mechanism can ultimately boost the superconducting $T_{\text{c}}$ of the entire system.

\subsection{Low energy electronic structure of Bi2223}

\begin{figure}[t!]
\includegraphics[width=18cm]{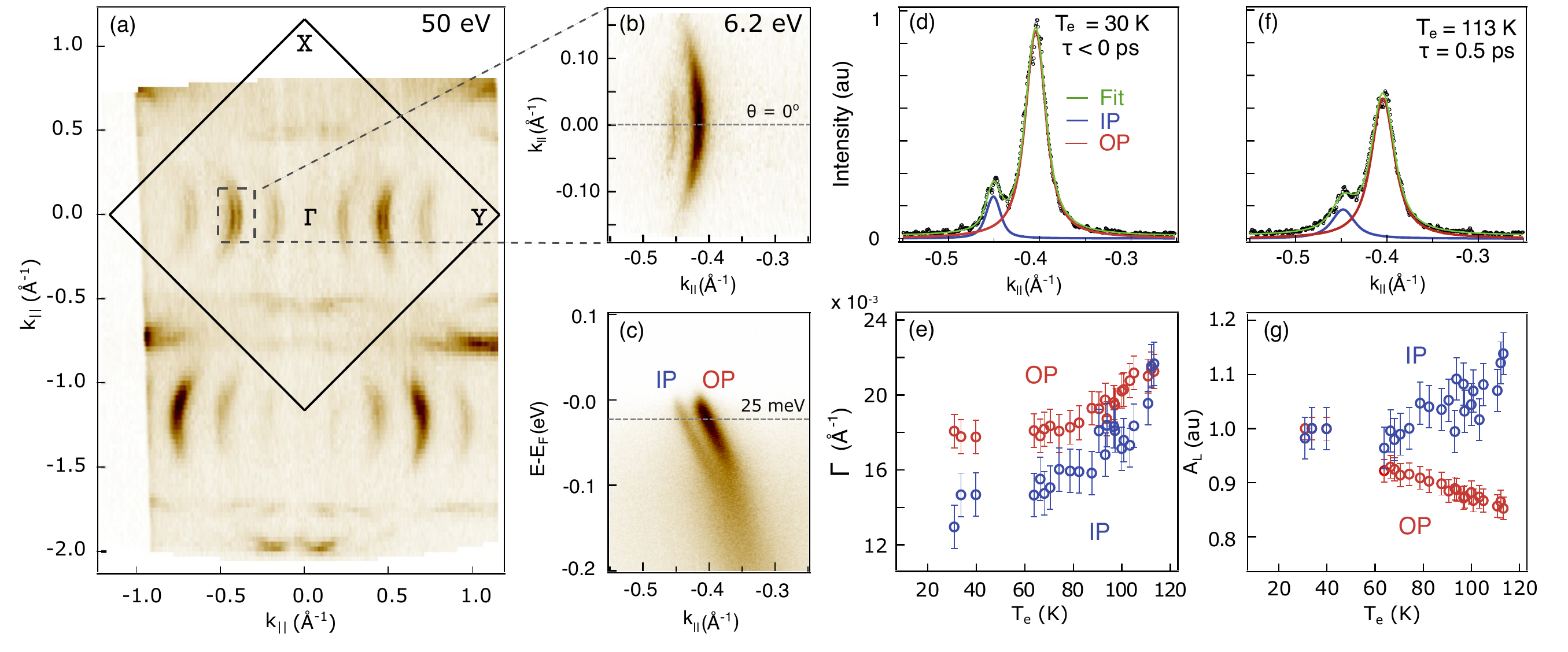}
\caption{{\bf Ultrafast scattering rates and spectral weight dynamics in the nodal dispersions of Bi2223.}  FS of the optimally doped trilayer cuprate Bi2223 ($T_{{\text c}}$~=~108~K) measured at $T =$ 15K using probe photon energies of 50~eV (a) and 6.2~eV (b). (c) Dispersion along the nodal $\Gamma Y$ direction at equilibrium, showing both the inner plane (IP) and outer plane (OP) bands crossing the Fermi level. The MDC at 25~meV as indicated by the dashed grey line in (c), is shown before (d) and after (f) excitation with a 1.55 eV pump, including a two-Voigt fit. (e) Lorentzian full width at half max ($\Gamma$) of the MDC peaks as a function of the effective electronic temperature T$_e$. (g) Temperature-dependent Lorentzian area A$_L$ normalized to equilibrium values (i.e. negative time delays corresponding to the lowest T$_e$), for both bands. The errorbars used are the largest systematic uncertainty given by the fitting procedure for each parameter, increased by an additional 50\% }
\label{Fig1}
\end{figure}

Bi2223 is the most extensively studied trilayer system by ARPES owing to its high critical temperature and natural cleavage plane~\cite{Feng2002,Sato2002,Muller2002}. The low-energy electronic states of Bi2223 originate from its stack of structurally distinct CuO$_{2}$ planes -- two symmetry-equivalent, heavily doped OP sandwiching a lightly doped IP, which is screened from inhomogeneities in the spacer layers. This structure leads to multiple FS with differing volumes, gaps, and kinks, all of which can be resolved with high-resolution ARPES~\cite{Ideta2010a, Ideta2013, Luo2023, Kunisada2017}. These previous works have also demonstrated hybridization between the Bogoliubov QP dispersions of the different FS in the superconducting state~\cite{Kunisada2017} and, for heavily overdoped Bi2223, even the presence of bonding-antibonding splitting of the bands originating from the two outer planes~\cite{Luo2023}. In Fig.~\ref{Fig1}(a) we show the FS of Bi2223 covering multiple Brillouin zones (BZ) measured at 15~K, displaying two FS sheets that are well separated along the nodal direction and are gapped out towards the zone boundary due to superconductivity. Assuming for simplicity a tetragonal BZ \cite{Smit2025}, we infer doping levels of 0.08 $\pm$ 0.02 (IP) and 0.25 $\pm$ 0.02 (OP) holes/Cu for these FS (see S.M.).

The intensity measured in an ARPES experiment,  $I({\bf k}, \omega)$, is directly proportional to the spectral function $A({\bf k},\omega)$~\cite{Damascelli2003, Damascelli2004, Sobota2020}, consisting of a coherent pole and an incoherent smooth part that have their relative weights determined by the QP residue $Z$. With our low-energy laser ARPES setup, we can investigate the single particle spectral function of the two distinct Fermi sheets in the nodal region with high momentum, energy, and time resolution~\cite{Dufresne2024}. 
We start by analyzing our nodal ARPES data via momentum distribution curves (MDC), which are slices through the data at fixed binding energy. We assume a linear bare dispersion $\epsilon_k = v_F(k-k_F)$ close to the Fermi level for both bands, and a self-energy that is not strongly momentum dependent perpendicular to the FS [$\Sigma (\omega,{\bf k}) \approx \Sigma (\omega)$]. This means the coherent parts of the spectral function can be very well approximated by a Lorentzian at fixed energy \cite{Smit2024} (for a full derivation of the link between fitting parameters and physical quantities see the S.M.). Here we will use the fact that the width of the Lorentzian MDC, $\Gamma(\omega)$, is proportional to the QP residue in momentum and the imaginary part of the self-energy via $\Gamma(\omega)= Z_{\bf k}\Sigma^{''}(\omega)$, while the spectral weight encoded by the area of the Lorentzian MDC, A$_L$, is only proportional to the QP residue as $A_L \propto Z_{\bf k}$. 

Fig.~\ref{Fig1}(d,f) show two MDC at 25 meV before and 0.5~ps after photoexcitation with an ultrafast 1.55~eV pump pulse. Included are two-Voigt fits where the Gaussian components are fixed to 0.009~\AA$^{-1}$ to account for the angular resolution. After converting the time delay to the effective electronic temperature T$_e$ (see. S.M. and Ref.~\onlinecite{Zonno2021b}), we can analyze the width and area (spectral weight) of the Lorentzian component [Fig.~\ref{Fig1}(e,g)] as a function of temperature for both under- and overdoped bands. In both cases we observe sharp peaks close the Fermi level, pointing to well-defined and coherent excitations at all electronic temperatures irrespective of the doping of each FS (the OP peaks are more intense than those of the IP at these photon energies \cite{Ideta2010a}, however this is not expected to influence their relative lineshape as a function of temperature). The OP show a modest momentum broadening upon increasing temperature, corresponding to a decreasing lifetime of the excitation of about $\sim$ 15~\% at T$_{el} \sim$ 113 K, together with a decrease of spectral weight. In contrast the IP -- which at low temperatures is slightly sharper than the OP, presumably owing to a lower degree of disorder  \cite{Mukuda2012} --  shows  a much stronger increase in its momentum width ($\sim$ 60~\% at T$_{el} \sim$ 113 K), and a marked increase in spectral weight, making it clear that the two FS exhibit a very distinct response to the pump excitations along the nodal direction.  While a loss of nodal coherent spectral weight with temperature has been observed both in single and bilayer Bi-based cuprates \cite{Zonno2021, Graf2011}, the coherent spectral weight in the correlated metal Sr$_2$RuO$_4$  was seen to increase with temperature~\cite{Hunter2023}. Theoretically this poses an interesting question as such behavior can strongly depend on the dominant mechanism suppressing QP coherence \cite{Skrlec2024}.

\subsection{Observation of layer-selective charge order by tr-ARPES and RXS}

\begin{figure}[t!]
\includegraphics[width=14cm]{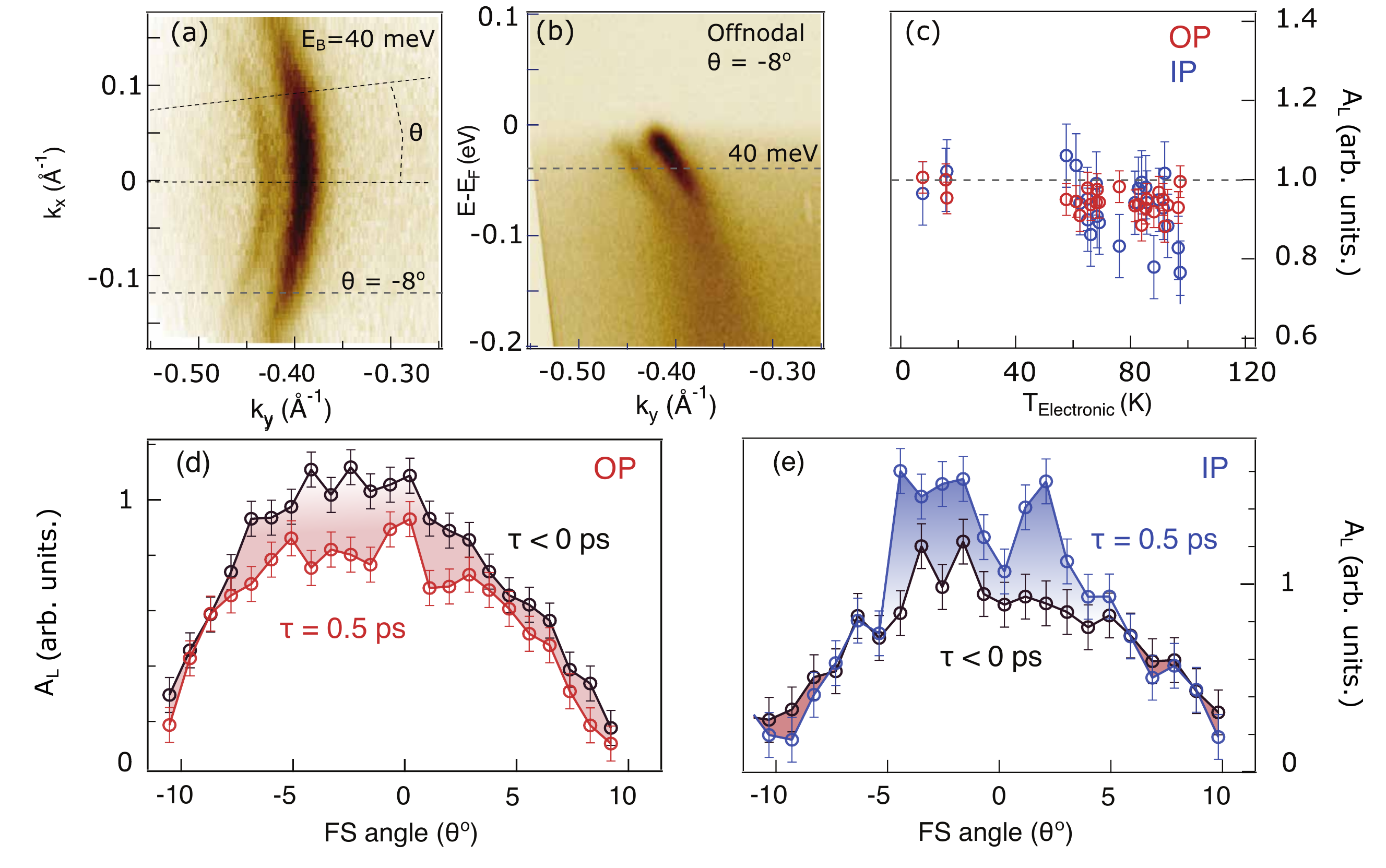}
\caption{{\bf Anisotropic momentum response of the spectral weight.}  (a) Equilibrium constant energy contour centered at $E_\textit{B}~=~40$~meV, integrated within a 20 meV window. (b) Off-nodal dispersion at $\theta = -8^{\circ}$ [grey line in (a)]. (c) Temperature dependence of the spectral weight at 40 meV [dashed black line in (b)] of the off-nodal dispersion, normalized to the low temperature value. (d,e) Spectral weight as function of FS angle $\theta$ at 40~meV for the OP and IP bands respectively, both before and after photo excitation; these curves are normalized to the nodal values before pumping (i.e. at negative delays).}
\label{Fig2}
\end{figure}

\begin{figure}[t!]
\includegraphics[width=15cm]{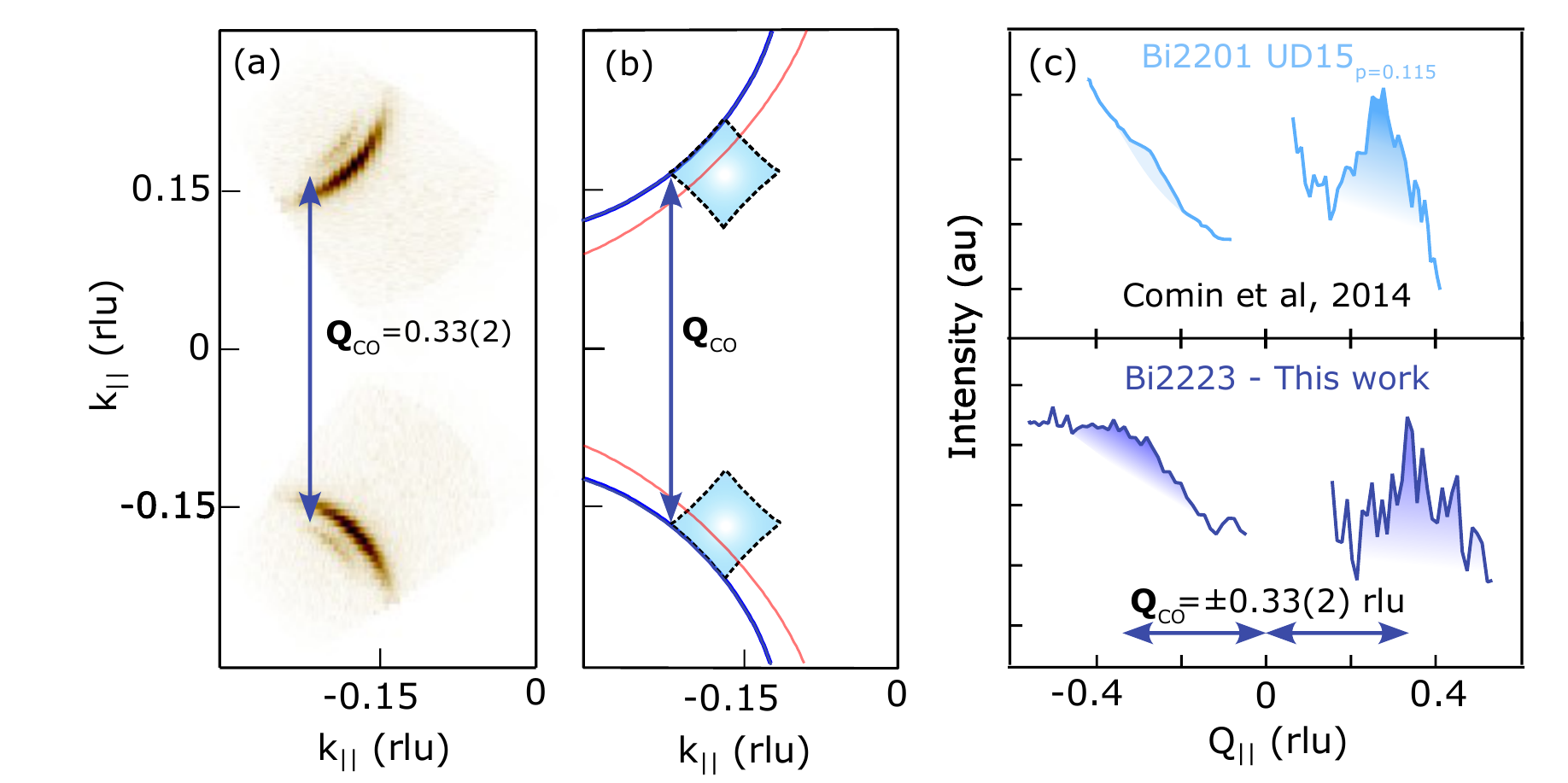}
\caption{{\bf Short range charge-order reconstructing the inner plane Fermi surface.} (a) Symmetrized FS in reciprocal lattice units (rlu), where the k-space loci on the IP that show the crossover in spectral weight dynamics [the $\pm6^{\circ}$ angular range from Fig. \ref{Fig2}(e)] are indicated with blue arrows. (b) Tight binding model for the Bi2223 FS (blue lines IP, red lines OP), together with a sketch of the reconstructed FS (black dotted lines) from a $Q~\simeq~\pm ~0.33$~r.l.u.  translational symmetry breaking.  (c) RXS at the Cu L-edge from Bi2223, confirming a $Q~\simeq~\pm ~0.33(2)$ r.l.u. CO wavevector, compared to results from heavily underdoped Bi2201 (Ref. \onlinecite{Comin2014b}).}
\label{Fig3}
\end{figure}

To better understand this discrepancy in the nodal dynamics, we perform a comprehensive analysis as a function of FS angle. Fig.~\ref{Fig2}(a) shows a constant energy contour of the ARPES intensity at 40~meV~$\pm$~10~meV (where ~$\pm$~10~meV refers to the 20 meV integration window), significantly below the superconducting gap energy and Fermi broadening in this range of momenta (see S.M.). In Fig.~\ref{Fig2}(b) we present the off-nodal dispersion at a FS angle of $\theta~=-8^{\circ}$, with the temperature-dependence of the spectral weight for both bands shown in panel (c), this time clearly displaying an analogous spectral weight reduction as function of electronic temperature, at variance with the dichotomy observed along the nodal direction in Fig. \ref{Fig1}(g). Using a similar 2-Voigt fitting procedure as before, we extract the spectral weight of both bands before and after photoexcitation around the BZ, with both contours exhibiting a maximum along the nodal direction [Fig.~\ref{Fig2}(d,e)]. While the OP spectral weight decreases upon pumping for all measured angles, the IP exhibits a remarkable crossover behavior: the observed increase of spectral weight at the node sharply changes to a decrease at, and beyond, FS angles of $\pm ~6^{\circ}$. Below we argue that, independently of a microscopic quantitative description of the observed spectral weight variation, the angular dependence is consistent with a broken translational symmetry along the Cu-O bond direction. In a single-particle picture, such a translational symmetry breaking would reconstruct the FS to a nodal electron pocket, as schematically shown in Fig.~\ref{Fig3}(b), with band gaps in the nodal region below the Fermi level where the crossing of the IP band and its back-folded replica is avoided.  However, in a strongly correlated system like our lightly doped IP, such a `nodal gap' at finite binding energy would rather present itself as a suppression of coherent spectral weight. Thus, the observed increase of spectral intensity with electronic temperature can be understood as the redistribution of previously incoherent spectral weight into the coherent poles of the spectral function; in other words, this would correspond to the filling-in rather than the closing of an ostensible hybridization gap below $E_{\text{F}}$. We note also that, since this analysis is performed at energy larger than the near-nodal superconducting gap itself, this phenomenology is not affected by the temperature evolution of the latter.

To verify our hypothesis of an incipient CO as revealed by tr-ARPES, we search for signs of translational symmetry breaking in diffraction. Fig. \ref{Fig3}(c) presents RXS data acquired at the Cu~$L_3$ resonance from Bi2223 samples at 17~K. These results reveal the presence of short-range CO, as indicated by broad peaks at both positive and negative momenta, $Q \simeq \pm 0.33(2)$~r.l.u.. These wavevectors match the $\pm 6^{\circ}$-range unconventional IP spectral weight dynamics, thus providing a natural explanation for the momentum dependence observed in Fig. \ref{Fig2}. While these RXS measurements establish the presence of bulk short-range CO at this wavevector, its assignment specifically to the IP follows from the layer-selective momentum dependence observed in the complementary tr-ARPES measurements. Comparing the RXS data to experiments performed on single layer Bi2201 samples at low doping  ~\cite{Comin2014b}, we note very comparable scattering signatures. Our Bi2223 has a larger wavevector compared to the Bi2201 ($Q = 0.33$ vs $0.26$), which can be attributed to the slightly higher doping of the single layer system compared to our IP ($p=0.115$ vs $p=0.08$). Bi2212 samples at even lower nominal doping have been shown to exhibit ordering wavevectors \cite{Boschini2021} closer to our measured value. More importantly, all these systems have similarly broad peaks, indicating the very short correlation length ($\xi_{\text{CO}} \sim 20 - 30$ \AA) of the CO. This is consistent with a lack of back-folded bands or a clean gap opening in either the MDC or EDC (energy distribution curves), and reminiscent of the short-range AFM order that has been observed to reconstruct the FS of the electron doped cuprates~\cite{He2019}; however, while reconstruction due to AFM order would also cause bandfolding in the nodal region, it would give rise to hybridization gaps above E$_{F}$, opposite of what we observe here. 

\subsection{Layer-dependent quasiparticle weight and pairing gap}

\begin{figure}[t!]
\includegraphics[width=15cm]{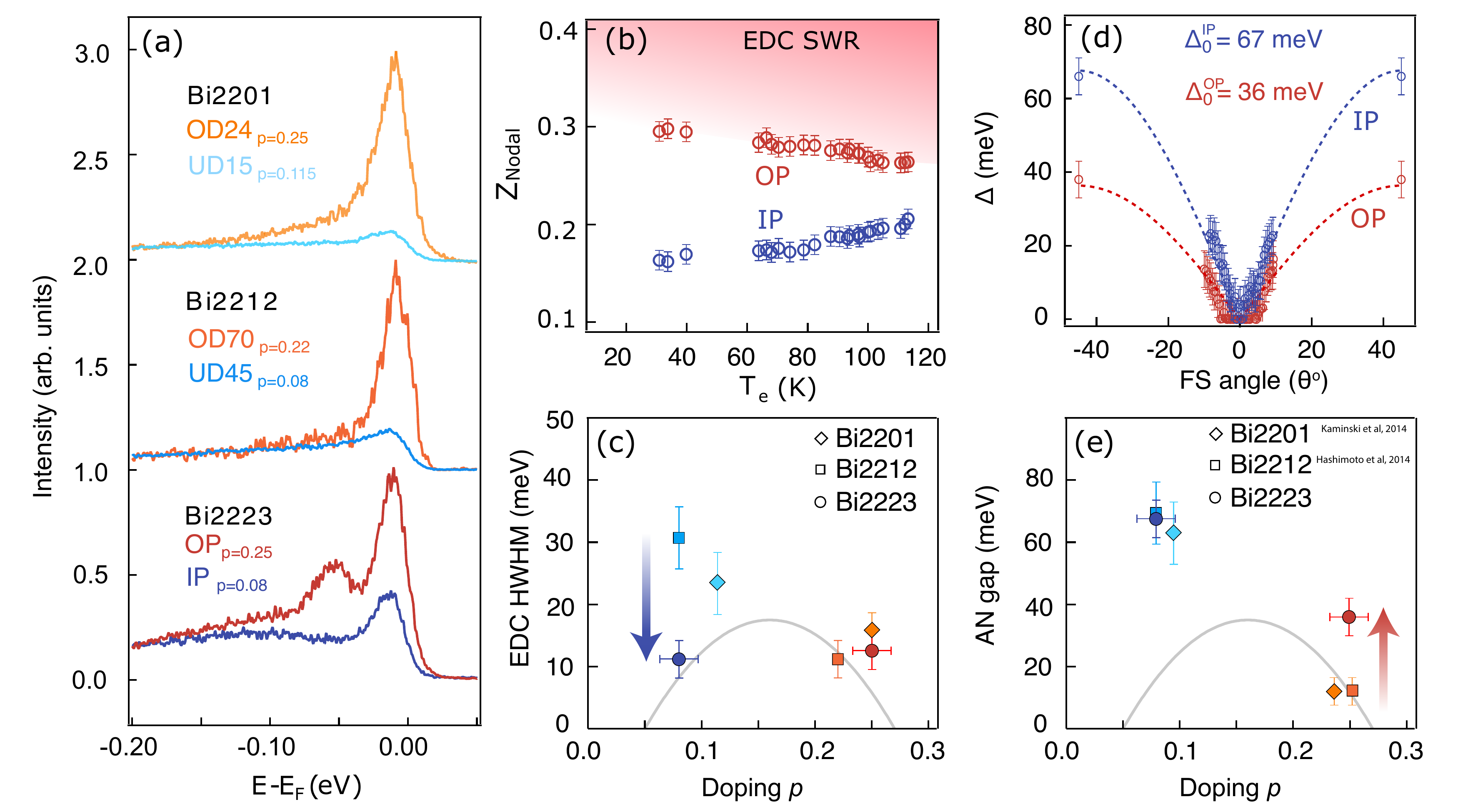}
\caption{{\bf Quasiparticle weight, scattering, and pairing.} (a) Energy distribution curves (EDC) at the nodal Fermi momentum $k_{\text{F}}$ for both the IP and OP bands, as well as for equivalently doped single (Bi2201) and bilayer (Bi2212) compounds. Each overdoped (OD) spectrum is normalized to the QP peak height, and the backgrounds of the underdoped (UD) spectra have been scaled to match. (b)  Temperature dependence of the QP residue $Z$ of the Bi2223 IP and OP nodal peaks, here defined as $Z = \frac{\int_{-0.025}^{\infty}I(k_F,\omega)d\omega}{{\int_{-\infty}^{\infty}I(k_F,\omega)}d\omega}$. As explained in the main text, the values for the OP (red markers) provide a lower bound of the true value indicated here by the gradient coloring. (c) Half-widths-at-half-maximum (HWHM) values of the nodal peaks shown in (a). (d) Superconducting gap along both IP and OP FS at 15 K together with a fit to $\Delta (\theta) = \Delta_0 \text{sin}(2 |\theta|)$; details on the SC gap extraction are given in the S.M. (e) Antinodal gap values of the three compounds together; the Bi2201 data is taken from Ref. \onlinecite{Kaminski2014}, and the Bi2212 from Ref. \onlinecite{Hashimoto2014}. The errorbars on these data points represent the uncertainty in digitization of the literature data, while the errorbars in (b-d) represent the largest systematic uncertainty given by the fitting procedure for each parameter, increased by an additional 50\%.}
\label{Fig4}
\end{figure}

Although the underdoped single and bilayer Bi-based cuprates also host CO correlations~\cite{Comin2014b, Boschini2021} similar to those observed on the IP of our trilayer Bi2223, the temperature evolution of the nodal spectral weight of the latter is opposite to what has been reported in single and bilayer materials~\cite{Graf2011,Zonno2021}. To obtain a more quantitative understanding of the distinct temperature evolutions of the IP and OP nodal spectra, we extract the QP strength $Z$ from our data following the procedure described in Ref. \onlinecite{Fournier2010}. In Fig. \ref{Fig4}(d) we plot $Z = \int_{-0.025}^{\infty}I(k_F,\omega)d\omega~ /~ {\int_{-\infty}^{\infty}I(k_F,\omega)}d\omega$. The integral in the numerator captures the spectral weight in the coherence peak, and the denominator integrates the entire EDC (see S.M. for more details about the procedure and choice of integration window). The heavily doped OP do have relatively more of their spectral weight in the coherent peak compared to the IP [Fig. \ref{Fig4}(a) bottom]. The OP EDC at $k_F$ also crosses the IP dispersion at a higher binding energy, which artificially adds spectral weight to the OP `incoherent' part; as a result, the extracted value for $Z_\text{OP}$ displayed in Fig.~\ref{Fig4}(b) should be considered as an effective lower bound of the true value. The difference in coherence between the planes, however, robustly decreases with increasing electronic temperature. This trend evidences an interlayer coupling, by which the IP gains coherence in the presence of the overdoped OP. Crucially, CO correlations compete with this proximity effect to suppress the QP coherence on the IP at low temperature. 
To further support the evidence for boosted coherence in the trilayer structure we present a systematic comparison of nodal spectra from the single-, bi-, and trilayer materials for both the under- and overdoped cases in Fig.~\ref{Fig4}(a). For all three compounds we show two nodal EDC: one for a lightly doped FS, and one for an overdoped FS. All three overdoped EDC show a strong, coherent peak at $E_{\text{F}}$, with a width in energy comparable to the instrumental resolution. In contrast, the electronic structure of underdoped Bi2201 and Bi2212 is characterized by an incoherent spectral function, with almost no discernible peak at $E_{\text{F}}$. Surprisingly, the underdoped IP of Bi2223 shows a much more prominent and coherent QP peak at the Fermi energy. This is further summarized in Fig. \ref{Fig4}(c), where we plot the half-width-at-half-max (HWHM) of the QP peaks in energy as a function of doping, showcasing in particular the equivalence of IP and OP lineshapes for Bi2223. 

To estimate the corresponding superconducting gaps in the near-nodal as well as antinodal region, we have complemented our laser-ARPES experiments with measurements performed with 50 eV photons at the Quantum Materials Spectroscopy Centre beamline at the Canadian Light Source (see S.M.). The superconducting gap extracted along both FS sheets is presented in Fig. \ref{Fig4}(d), with the IP having a maximum value at the antinode $\Delta_{0}^{\text{IP}} = $ 67 meV, which is approximately twice the OP value $\Delta_{0}^{\text{OP}} = $ 36 meV. Interestingly however, this OP value  is still more than two times larger than what is reported for similarly overdoped bilayer Bi2212 \cite{Ideta2010a, Hashimoto2014}, as shown in Fig. \ref{Fig4}(e).  The comparison of the spectra and gaps from the three compounds supports the picture that the proximity of the IP and OP is responsible for the sharp spectral features, unusual temperature dependence of the coherent spectral weight, and exceptionally large superconducting gaps. 

\subsection{Layer-resolved correlations and superconductivity: theoretical perspective}

In order to gain more insight about the microscopic mechanisms behind the observed electronic phenomenology, we performed theoretical calculations of a model three-layer cuprate where each layer is described by a single-band Hubbard model and the planes are coupled by a hopping term (details and parameters can be found in the methods section).
The model is solved using a cluster rotationally-invariant slave-boson method (CRISB) ~\cite{Lechermann2007}, which allows us to compare the normal and superconducting state ~\cite{Isidori2009}, and gives direct access to the layer-selective properties of the material.

\begin{figure}[t!]
\includegraphics[width=11.5cm,angle=0]{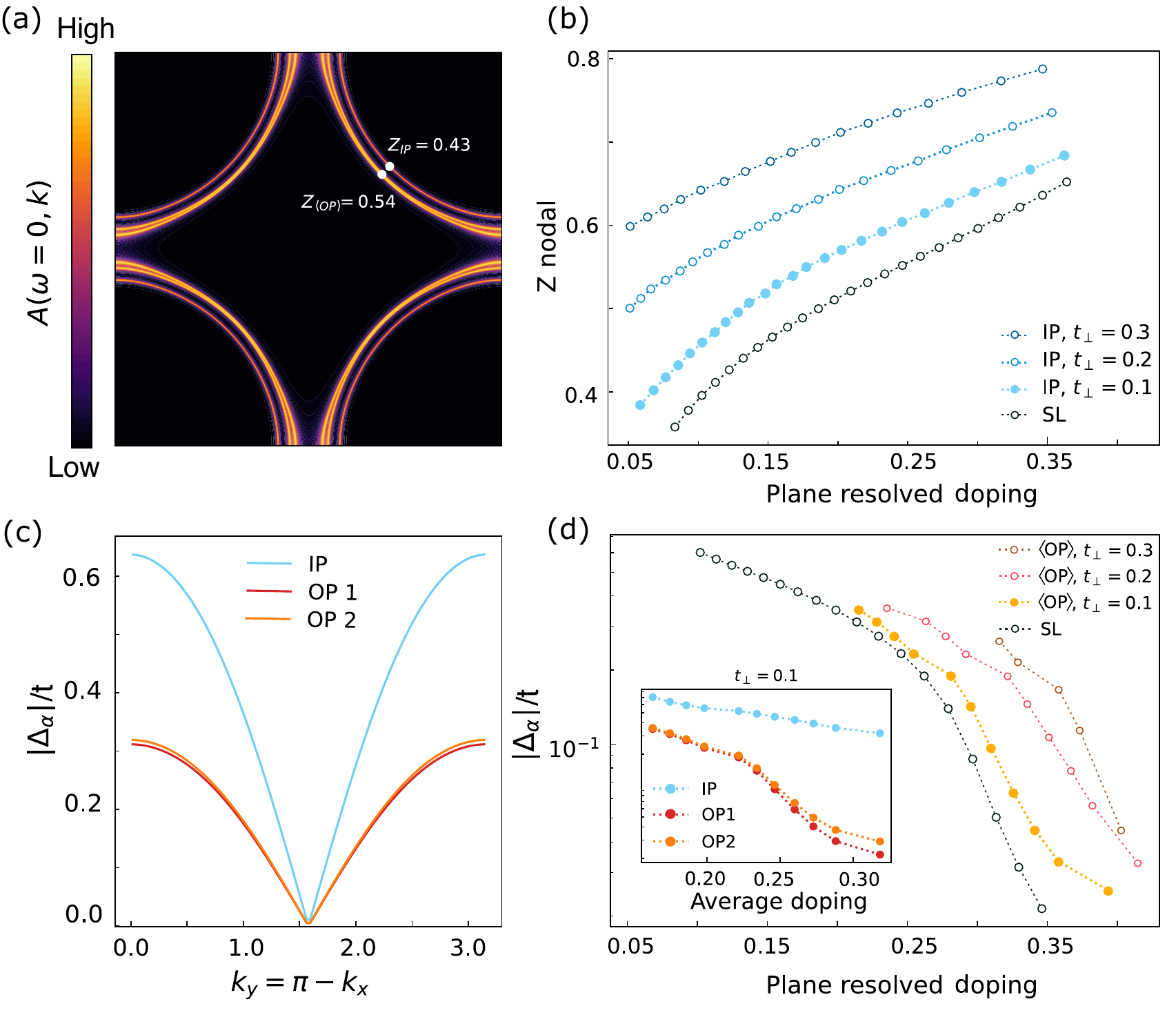}
\caption{ {\bf Theoretical analysis for a model three-layer system.} (a) Fermi surface of the IP and OP in the normal state, obtained from the spectral function $A(\omega,\textbf{k})$ at zero frequency for $t_\perp = 0.1$, with the central layer having a lower local energy $\epsilon_{IP} = -1.0t$ compared to the outer layers. The layer-resolved dopings are $p_{\text{OP1}} = 0.23$, $p_{\text{OP2}} = 0.28$ (leading to an average OP doping $p_{\text{\textlangle OP\textrangle}} = 0.25$), and $p_{\text{IP}} = 0.09$. The QP residue $Z$ is shown at the nodal points. (b) Nodal QP residue Z for the IP as a function of doping, for three different values of the interlayer hopping, compared to a single-layer model (labeled SL). (c) Superconducting spectral gap $|\Delta_\alpha|$ along the direction $k_y=\pi-k_x$ calculated for $t_\perp = 0.1$. (d) Average OP superconducting gap $|\Delta_\alpha|$ at the antinodes as function of doping for three different values of the interlayer hopping, compared to a single-layer model. The inset shows the SC gap for the three FS sheets as function of the average doping of the trilayer for $t_\perp = 0.1$.}
\label{Fig5}
\end{figure}

The results are summarized in Fig. \ref{Fig5}. Panel (a) shows the computed normal state spectral function at the Fermi energy and the corresponding quasiparticle weights $Z$ for the IP and OP, for a global doping $p = 0.19$ and using an interlayer hopping $t_\perp = 0.1$. This shows that our calculations can correctly reproduce the layer-dependent electronic properties in the normal state, which we associate with the different doping of the planes, matching expectations from first-principal calculations \cite{Bacq2025}. We resolve three Fermi surfaces that we call IP, OP1, and OP2. Note that OP1 and OP2 here refer to the two bonding and antibonding bands that mainly originate from the outer planes, rather than to two bands corresponding each to one of the two outer planes. The IP has a smaller hole doping $p_\text{IP} = 0.09$, while the two OP are more overdoped ($p_\text{OP1} =0.23$ and $p_\text{OP2} = 0.28$), as a result of the lower on-site energy of the inner layer dictated by the local crystal-field symmetry (see Methods). While our CRISB calculations overestimate the absolute value of $Z$ for IP and OP, the layer-resolved trend we find (higher Z on the OP) implies that also the correlation properties of the IP and OP are, respectively, similar to those of underdoped and overdoped single-layer cuprates. Previous studies show that improving the approximation generally reduces all Z with respect to CRISB results, while preserving their relative hierarchy, so we expect the same qualitative trend to hold here \cite{Potthoff2001,Lee2023}. In Fig.~\ref{Fig5}(b) we show the nodal QP residue Z of the IP, for a range of plane-resolved dopings and interlayer hopping values. Importantly, compared to a single layer model with the same in-plane hopping and coulomb U, we find that the IP has a significantly larger $Z$. This effect is ubiquitously present for all parameters we have explored, and grows with increasing $t_{\perp}$.

The interpretation of the ARPES data in terms of an underdoped IP coupled to overdoped OP is further supported by the results for the superconducting phase presented in Fig. ~\ref{Fig5}(c,d).  Panel (c) shows the $d$-wave superconducting gap, with an amplitude which is significantly larger for the IP than for the OP, in agreement with the experimental data of Fig. \ref{Fig4}(e). In Fig. ~\ref{Fig5}(d) we show the evolution of the OP superconducting gap at the antinode as a function of doping and interlayer hopping. As we do not resolve the OP splitting in our experiments, we choose here to plot the average value of the two OP bands ($\langle OP\rangle$). Significantly, at high doping the OP have a much larger gap than a single-layer. The inset shows that the gaps on each individual layer are a monotonically decreasing function of the global doping, with the OP1 and OP2 values being very close across the doping range. At any value of the average global doping, the IP is more underdoped than the OP, and thus always has a larger gap as is expected in any framework where strong correlations and/or spin fluctuations are key for superconductivity.  We verified that the doping differentiation and the concomitant enhancement of both nodal Z and the superconducting gap are generic by comparing different choices of model parameters (see S.M.); this demonstrates conclusively that the effect is robust and is thus not an artifact of arbitrary parameter tuning, and can be expected to further increase upon treating more accurately the many-body interactions between the layers.

\subsection{Discussion}

The data presented in this work contain multiple intriguing observations regarding the low-energy electronic structure of the trilayer cuprate Bi2223. Both the experimental data and theoretical calculations indicate that this system consists of interacting planes of differing doping, where the strongly underdoped IP has unexpectedly high QP weight, and the OP a relatively large gap. 

The momentum-dependent spectral weight dynamics along the IP FS establishes the first observation of CO in the Bi-based cuprates by ARPES. In complementary RXS experiments, we have confirmed the presence of this order, along with its wavevector and short correlation length. Of note is that magnetotransport signatures of similar layer-dependent charge and spin-order have been observed in heavily underdoped trilayer Hg1223, where the charge-order occurs on outer CuO$_2$ planes with a doping very close to our IP \cite{Oliviero2022}. While the effects of such a charge order on the temperature dependent QP weight remains to be theoretically investigated, the corresponding order parameter can fluctuate, which is generally detrimental to coherent QP excitations. These fluctuations enter in the incoherent part of the spectral function, and are therefore expected to reduce $Z$. Upon increasing temperature, the fluctuating short-correlation length CO on the IP gets suppressed, and the QP residue of the IP gets supported towards the OP value. Interestingly, the OP bands remain unaffected by this order, indicating that significant breaking of translational symmetry occurs only on the IP. The lack of equivalent spectral weight signatures of CO in UD single and bilayer systems \cite{Zonno2021} indicates that the enhanced IP coherence inherited from the OP is critical for these subtle effects to be noticeable.

The IP has almost all the hallmarks of a strongly underdoped CuO$_2$ plane. It has a small Fermi volume of approximately $ 8 \% $ doping, a CO instability at low temperatures, and a significantly larger gap compared to the OP indicating stronger pairing interactions [shown in Fig. \ref{Fig4}(e) and the S.M.]. However, the large and qualitatively similar $Z$ between IP and OP, as determined from the sharp MDC peaks and coherent EDC spectra presented in Figs.~\ref{Fig1} and ~\ref{Fig4}, respectively, stand in clear contradiction to the expectation of near-zero QP residue in underdoped cuprates. Except in multilayer compounds with more than two CuO$_2$ layers per unit cell \cite{Kunisada2020, Kurokawa2023}, negligible quasiparticle weight at low doping is a ubiquitous feature in the cuprates: it has been shown in YBa$_2$Cu$_3$O$_{6+x}$ both by tracking of the nodal bilayer splitting and the spectral weight ratio \cite{Fournier2010}; in the Bi-based cuprates (both single- and bilayer), it has been demonstrated before this work at the antinodes ~\cite{Chen2019,Berben2022} as well as the nodes \cite{Reber2019, Smit2024}, as a consequence of the power-law-liquid phenomenology. Our finding of an enhanced quasiparticle coherence on the IP, concomitant with an increase of the pairing gap on the OP, strongly challenges the notion that the electronic excitations within the IP and OP of Bi2223 are purely representative of those found in `model' under- and overdoped cuprates.  

It is important to realize that the chemical environment of the IP is distinct from the OP, and from the CuO$_2$ planes found in the single and bilayer systems. Evidence points towards differing levels of disorder in IPs and OPs  \cite{Mukuda2012}, consistent with the small differences in spectral width between the IP and OP bands in Fig. \ref{Fig1}(e). It has even been suggested that the high coherence of the IP compared to single and bilayer systems could be the result of a `screening' from out-of-plane disorder by the OP \cite{Kunisada2020}. Such a scenario would imply that the broad nodal spectra in heavily underdoped single and bilayer cuprates are mainly due to such disorder, instead of predominantly being a consequence of increased correlations at low doping. A clear illustration against this line of argumentation comes from single-layer La$_{2-x}$Sr$_x$CuO$_4$ and Na$_{2-x}$Ca$_x$CuO$_2$Cl$_2$: both materials are exceptionally clean in their undoped x=0 insulating phase, yet in this pristine limit the ARPES lineshape remains extremely broad (up to ~500 meV) and only evolves into a sharp and coherent quasiparticle peak upon increasing cation substitution well beyond x = 0.1 -- and with it both carrier doping and disorder \cite{Shen2004}.
In addition, a scenario where screening of disorder is the driving force behind the enhanced IP coherence would not explain the simultaneously enhanced pairing on the OP as compared to single and bilayer systems. 

In conclusion, the picture emerging from our combined experimental and theoretical investigation is that of the coexistence between strong-coupling superconductivity in the underdoped IP and weak-coupling superconductivity in the overdoped OP, where proximity between the layers increases the coherence of the IP as well as the pairing of the OP. This is similar to what is found in multiorbital models describing iron-based superconductors, where the doping happens in an orbital-selective way, leading in turn to orbital-dependent correlations that are fundamental for the normal-state properties of these materials ~\cite{deMedici2014}. We also note that a similar proximity between strong- and weak-coupling layers has also been shown to give rise to an enhancement of the critical temperature in a model of heterostructures with $s$-wave pairing \cite{Mazza2021}, realizing the scenario originally proposed in Refs. \onlinecite{Kivelson2002,Berg2008}.

\subsection*{Methods}
\subsubsection{ARPES}
Equilibrium ARPES measurements were performed at the Quantum Materials Spectroscopy Centre (QMSC) beamline of the Canadian Lightsource using 50 eV vertically polarized photons and a Scienta R4000 hemispherical analyzer, with a total energy resolution of 15 meV. All the samples were aligned along the $\Gamma Y$ direction to avoid contamination of the bands due to the supermodulation. The time-resolved ARPES experiments were conducted at the UBC-Moore Center for Ultrafast Quantum Matter using a Scienta DA-30L hemispherical analyzer and vertically polarized pump (1.55 eV), and probe (6.2 eV) photons. Different photon energies modulate the relative intensity between the IP and OP bands \cite{Ideta2010a}, but this is not expected to have any impact on their lineshape as function of temperature or pump probe delay. The probe and pump beams were focused onto the sample with a 45$^{\circ}$ angle of incidence, with a FWHM beamprofile of 50 $\mu$m and 100 $\mu$m respectively. The energy and
temporal resolutions of the system are 19 meV and 290 fs. The penetration depth of the pump-beam is much larger (160 nm) than the escape depths of the measured photoelectrons, so we assume the outer and inner CuO$_2$ planes to be homogeneously illuminated. In the time-resolved data in Fig. \ref{Fig2} there is a noticeable dip in the IP spectral weight around a FS angle of 0 degree. We suspect this dip is due to an analyser inhomogeneity/electron optics artifact that showed up when performing a Fermi surface mapping using electrostatic deflectors. In both ARPES systems the samples were cleaved below 15 K, at a base pressure lower than 7 x $10^{-11}$ mbar. 
The Bi2201 UD15K data shown in Fig \Ref{Fig4}(a,c) was taken on a crystal from the same growth as the data from Ref. \onlinecite{Comin2014b} that was reproduced in Fig. \ref{Fig2}(c). We use the same nominal doping (p=0.115) for these datapoints. For the datapoints in Fig.\ref{Fig4}(e) that are adapted from Refs \onlinecite{Kaminski2014,Hashimoto2014}, the Pressland-formula was used to calculate a doping-value from the critical temperature \cite{Berben2022}.  

\subsubsection{RXS}

RXS experiments at the Cu L$_3$ edge (932 eV) were carried out at beamline 13-3 of the SSRL \cite{Kuo2025}. The samples were cleaved in air to expose the (0, 0, 1) surface before RXS measurements, and mounted on an in-vacuum 4-circle diffractometer. The CDW peak profiles along the in-plane momentum direction H were measured by performing scans of the sample angle $\theta$ with respect to the incident X-ray beam, while the detector was fixed at 137.5[165]$^{\circ}$ for the positive[negative] H-side. The samples were cooled using an open-cycle helium cryostat with a base temperature of 17 K. In order to subtract fluorescence backgrounds accurately, the scans were repeated with the detector moved out of the scattering plane.

\subsection{Theory}

We consider a three-layer single-band Hubbard model with nearest- and next-nearest-neighbour hopping $t$ (that we use as the energy unit) and $t' = -0.35t$ on the planes, and an interlayer tunneling $t_\perp = - 0.1 t$. The local Coulomb repulsion is $U=11.1 t$. The central layer has a lower local energy $\epsilon_\text{IP} = -1t$ while the two outer layers have the same energy $\epsilon_\text{OP} =0$. The model is solved using the cluster-rotationally invariant slave-boson method (CRISB)\cite{Lechermann2007,Isidori2009}, which successfully reproduces most of the experimental features at an affordable computational cost.

We note that our approach works at zero temperature and is based on a quantum embedding of an elementary unit of a 2$\times 2$ plaquette within each layer, while we keep the individual layers free to have different physical properties. We allow for $d$-wave superconductivity, while we cannot describe the charge-ordering patterns observed in the cuprates. This approach allows for a fast a thorough investigation of the large parameter space, but it does not aim at a quantitative description of the experimental results.

\subsection*{Data availability}

All data are processed by Igor Pro 9.0.5 software. All data needed to evaluate the conclusions in the paper are available within the article and its Supplementary Information files. All data generated during the current study are available from the corresponding author upon request.

\subsection*{Code availability}

The code generated during the current study is available from the authors on request.

\subsection{Acknowledgements}

We thank S.A. Kivelson, B. Ramshaw and Y.Y. Peng for helpful discussions.

\subsection*{Funding statement}

This research was undertaken thanks in part to funding from the Max Planck–UBC–UTokyo Centre for Quantum Materials and the Canada First Research Excellence Fund, Quantum Materials and Future Technologies. This project is also funded by the Natural Sciences and Engineering Research Council of Canada (NSERC), Canada Foundation for Innovation(CFI); the Department of National Defence (DND); the British Columbia Knowledge Development Fund (BCKDF); the Mitacs Accelerate Program; the QuantEmX Program of the Institute for Complex Adaptive Matter; the Gordon and Betty Moore Foundation’s EPiQS Initiative, Grant GBMF4779 to A.D. and D.J.J.; the Canada Research Chairs Program (A.D.); and the CIFAR Quantum Materials Program (A.D.). Use of the Canadian Light Source (Quantum Materials Spectroscopy Centre), a national research facility of the University of Saskatchewan, is supported by  CFI, NSERC, the National Research Council, the Canadian Institutes of Health Research, the Government of Saskatchewan and the University of Saskatchewan. The RXS experiments were carried out at the SSRL (beamline 13-3), SLAC National Accelerator Laboratory, supported by the U.S. Department of Energy, Office of Science, Office of Basic Energy Sciences under Contract No. DE-AC02-76SF00515. S.S. acknowledges support by the Netherlands Organisation for Scientific Research (NWO 019.223EN.014, Rubicon 2022-3). M.C., E. Z., G. B. and S. G. acknowledge support of MUR via PRIN 2020 (Prot. 2020JLZ52N 002) program, and by the European Union - NextGenerationEU through PRIN 2022 (Prot. 20228YCYY7), MUR PNRR Project No. PE0000023-NQSTI, and No. CN00000013-ICSC.

\subsection*{Author contributions}

S.S., M.B., A.D conceived the investigation. S.S performed the tr-ARPES measurements with assistance from M.B., P.M., S.K.Y.D., C.T.S., J.D., and S.Z.. M.M., G.L., J.D., S.Z., and D.J.J. were responsible for the operation and maintenance of the UBC tr-ARPES set-up. S.S., M.B., P.M., C.T.S., C.A.Y., and V.Z. performed the ARPES experiments at CLS with assistance from S.Go., P.D, and M.Z.. The RXS experiments at SSRL were performed by H.L., C-.T.K., and J-S.L.. Samples were grown and characterized  by D.S., S.I., H.E., and B.K.. Data-interpretation was the responsibility of S.S., M.B., N.H., I.E., and A.D.. 
E.Z., G.B., S.Gi. and M.C. designed and performed the theoretical calculations using CRISB.
All authors discussed the underlying physics and contributed to the manuscript. A.D. was responsible for the overall direction, planning, and management of the project.

\subsection*{Competing interests}

The authors declare no competing interests.

\newpage

\end{document}